\documentclass[12pt]{article}

\pdfoutput=1
\usepackage{arxiv}
\usepackage[utf8]{inputenc} 
\usepackage[T1]{fontenc}    
\usepackage{hyperref}       
\usepackage{url}            
\usepackage{booktabs}       
\usepackage{amsfonts}       
\usepackage{nicefrac}       
\usepackage{microtype}      
\usepackage{lipsum}		
\usepackage{graphicx}
\usepackage{natbib}
\usepackage{doi}
\usepackage{setspace}
\title{Identifying Undercompensated Groups Defined by Multiple Attributes in Risk Adjustment}

\author{ Anna Zink  \\
	Harvard University \\
    Cambridge, MA 02138 \\
	\texttt{azink@g.harvard.edu} \\
	\And
	Sherri Rose   \\
	Stanford University\\
	Stanford, CA 94305 \\
	\texttt{sherrirose@stanford.edu} 
}

\renewcommand{\shorttitle}{Identifying Undercompensated Groups}

\hypersetup{
pdftitle={A template for the arxiv style},
pdfsubject={q-bio.NC, q-bio.QM},
pdfauthor={Anna Zink, Sherri Rose},
pdfkeywords={Risk adjustment, Subgroups, Random Forests},
}

\begin{document}
\maketitle

\begin{abstract}
Risk adjustment in health care aims to redistribute payments to insurers based on costs. However, risk adjustment formulas are known to underestimate costs for some groups of patients. This undercompensation makes these groups unprofitable to insurers and creates incentives for insurers to discriminate.  We develop a machine learning method for ``group importance'' to identify unprofitable groups defined by multiple attributes, improving on the arbitrary nature of existing evaluations. This procedure was designed to evaluate the risk adjustment formulas used in the U.S. health insurance Marketplaces as well as Medicare. We find that a number of previously unidentified groups with multiple chronic conditions are undercompensated in the Marketplaces risk adjustment formula, while groups without chronic conditions tend to be overcompensated in the Marketplaces. The magnitude of undercompensation when defining groups with multiple attributes is larger than with single attributes. No complex groups were found to be consistently under- or overcompensated in the Medicare risk adjustment formula. Our work provides policy makers with new information on potential targets of discrimination in the health care system and a path towards more equitable health coverage.
\end{abstract}

\keywords{Risk adjustment \and Subgroups \and Random Forests}

\section{Introduction}
Risk adjustment is an essential tool in regulated health insurance markets. It redistributes health plan payments to insurers with higher cost patients, aiming to decrease the relationship between health and profits to ensure that sicker individuals are not discriminated against by insurers \citep{mcguire_risk_2018}. Risk adjustment for health plan payments has been successful in reducing selection incentives for many individuals, but incentives to discriminate still exist for groups of individuals whose costs are underpredicted by the risk adjustment formula, including those with mental health and substance use disorders \citep{mcguire_assessing_2014, montz_risk-adjustment_2016}.

The ability of insurers to discriminate against some groups of patients has been curbed since the passage of the Affordable Care Act (ACA) in 2010, which prevents insurers from refusing enrollment or changing premium prices based on enrollee health. However, insurers are able to attract more profitable enrollees or discriminate against less profitable enrollees through advertising as well as benefit design choices. Changes to benefit design include which providers comprise their provider networks and which drugs are placed in higher cost tiers or subject to drug formulary management tools \citep{shepard_hospital_2016, lavetti_strategic_2018, rose_computational_2017, carey_technological_2017, aizawa_advertising_2018, geruso_screening_2019}. Recently, machine learning methods have been used to identify individual conditions that are underpaid by the risk adjustment formula as well as drugs predictive of unprofitability, but these studies did not focus on groups defined by more than one attribute\citep{rose_computational_2017, rose_robust_2018}. In Germany and the Netherlands, researchers have incorporated interactions of variables into the risk adjustment formula based on the partitioning of a single regression tree predicting the residual, but only considered groups defined within the current risk adjustment formula \citep{buchner_regression_2017, van_veen_exploring_2018}. To date, no systematic method exists for identifying groups at risk for discrimination in the health care payment system, or more broadly in the algorithmic fairness literature, especially groups defined with a more complex set of attributes. Even groups defined by a single sensitive attribute are typically pre-specified \citep{chouldechova_fairer_2017}.   

In this paper, we present the first data-driven method for identifying undercompensated groups in health plan payment risk adjustment that are defined by multiple attributes. Extending the concept of variable importance for single attributes or predefined groups, we construct a new measure of ``group importance'' in the random forests algorithm that relies on multiple attributes \citep{gromping_variable_2009, louppe_understanding_2013, wehenkel_random_2018, fisher_all_2019}. Groups are defined based on the demographic and clinical categories susceptible to benefit redesign, namely chronic conditions, age, and sex. We designed our method for the purposes of studying the risk adjustment formulas used in the individual health insurance Marketplaces created by the ACA as well as Medicare. Our newly proposed method for identifying undercompensated groups overcomes the arbitrary nature of existing evaluations of risk adjustment, providing policy makers a tool for uncovering incentives for selection that persist in insurance markets and a path towards more equitable health coverage.

\section{Methods}
\label{sec:methods}

\subsection{Defining Groups}

Our primary objective is to identify complex groups vulnerable to discrimination in the health insurance market through benefit design. Therefore, in defining groups, we consider what factors are necessary for this type of discrimination. To start, the groups must be actionable for insurers, meaning the insurers must know which groups are profitable or unprofitable and be able to act upon this information via the design of provider networks or drug formularies \citep{ellis_predictability_2007}. The losses or gains caused by the group must also be large enough to substantially impact insurers' revenue. There are a number of factors that could lead to this type of large impact for insurers, including the size of the group, the size of the undercompensation of the group, and the persistence of the undercompensation over time \citep{mcguire_risk_2018}. For these reasons, we focus on identifying groups defined by prevalent chronic conditions that require specialist care or drug treatments. In addition to chronic conditions, we consider whether age and documented sex affect the size of undercompensation.

\subsection{Notation}

The canonical risk adjustment formula estimates individual annual spending $Y$ using a vector of input variables $X$ that contains $j$ demographic variables, $D=\{D_1,D_2,\dots,D_j\}$ and $k$ health condition categories, $H=\{H_1,H_2,\dots,H_k\}$ or $X=\{D,H\}$. The risk adjustment formula is some function $f$ mapping $X$ to $Y$,$ f\left(X\right)=Y$. The residual of the risk adjustment formula, $R=\hat{Y}-Y$, measures the under- or overprediction for each individual, where $\hat{Y}$ is predicted spending. We are interested in the set of groups $G$ predictive of the residual $R$. To define the groups, we select $s$ indicator variables $I=\{l_1,I_2,\dots, I_s\}$ that will comprise the components of the groups in $G$: a set of chronic conditions, sex, and age categories. A group in $G$ is defined as any combination of components in $I$. This allows us to consider various levels of complex groups. For example, older women with cancer and mental health disorders would be one group and women with mental health disorders another. We examine only binary variables; relaxing this restriction would increase the number of groups substantially. 

\subsection{Estimation}

Insurers can adjust their premiums at defined market levels to account for differences in costs in different geographic regions \citep{cms_marketplaces}. Because we are interested in undercompensated groups within markets, we want to adjust for market-level differences in spending. Therefore, we start by estimating a slightly modified risk adjustment formula using ordinary least squares regression: 

$$Y=\beta X+\textnormal{Market}+\epsilon,$$

\noindent where, beyond $X$, we additionally control for geographic market, Market, to account for routine premium adjustments made at the market-level. 

To find the groups in $G$ most predictive of the residual $R$, we predict $R$ with the group component variables $I$ using the random forests algorithm. The random forests algorithm grows many decision trees to predict $R$. Each tree uses a bootstrapped sample from the original data to decrease the variance. This is often referred to as bagging, but unlike other forms of bagging, the random forests algorithm also selects a subset of variables to consider at each split point in the tree so the same highly correlated variables don't appear in every single tree \citep{breiman_random_2001}. 

Tree-based methods are an intuitive way of forming groups because each decision tree splits the sample into a set of mutually exclusive groups. These group are defined based on the intersections of the input variables used for predicting the outcome. At each split in the tree, a variable is selected that best partitions the variable space in order to reduce variation within each new node. By using the component variables $I$ to predict $R$, we naturally create groups in $G$ that are interactions of the component variables in $I$ used to split the tree. For each tree, we find the set of groups that are formed in the terminal nodes. We aggregate this information across all trees, recording the number of times a group appears and its mean estimated residual $R$ across trees and years. See Figure 1 for an illustration of this process. 

\begin{figure}
\centering
\includegraphics[scale=1]{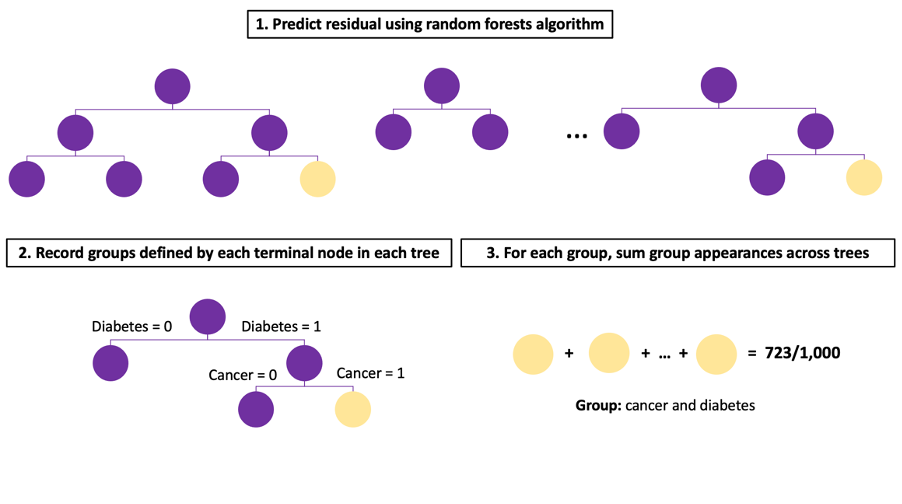}
\caption{Illustration of Random Forests for Group Importance}
\label{fig:fig1}
\end{figure}

The hyperparameters of the random forests algorithm are important for defining group criteria (in addition to algorithmic performance). Through hyperparameter specification, we set the minimize size of groups by requiring a minimum size for terminal nodes in the decision tree. We also control the complexity of the groups by limiting the depth of the tree and the number of terminal nodes. The deeper the tree grows, the more attributes the groups will contain. Recall that typically only a subset of randomly selected variables is considered at each split. By selecting a smaller number of variables at each split, we can decorrelate not only variables across trees but also groups, ensuring that the same groups are not favored by every tree. And finally, the more trees we grow, the more stable our results will be, but this can be computationally costly and after a certain number of trees only trivially improves stability \citep{probst_tune_2017}. 

\section{Marketplaces Risk Adjustment}

In 2014, the Department of Health and Human Services developed a risk adjustment formula for individual and small group markets to redistribute insurer payments in the Marketplaces newly established under the ACA \citep{kautter_hhs-hcc_2014}. Plans estimated to have lower risk enrollees subsidize plans with higher risk enrollees through a budget neutral risk adjustment transfer program regulated by CMS \citep{semanskee_explaining_2016}. Premiums are set within market rating areas defined by counties or Metropolitan Statistical Areas (MSAs) depending on the state \citep{cms_marketplaces}. The Marketplaces risk adjustment formula is built using the IBM MarketScan Commercial Database, which contains claims on over 23 million individuals receiving employer-sponsored private health insurance in the U.S. \citep{ibm_marketscan_2020}.

We estimated individual annual spending using the 2019 Marketplaces risk adjustment formula inputs adding geographic fixed effects at the MSA level using the following specification: 

$$Y= \textnormal{Age} \times \textnormal{Sex}+\textnormal{HCC}+\textnormal{MSA}+\epsilon,$$

where Age $\times$ Sex are ten age and sex categories and HCC are a set of Health Condition Categories mapped from diagnosis codes. We used MSAs as a proxy for premium rating areas in all states because we did not have county information for states. 

\subsection{Data}

To evaluate the Marketplaces risk adjustment formula, we sampled adults (aged 21-64) insured for a full year in either 2016, 2017, or 2018 from the Truven data. We calculated individual annual spending as the total spending across the individual's outpatient, inpatient, and carrier claims in the given year. The traditional risk adjustment formula predicts insurer spending, but for simplicity we used total spending, which included patient cost-sharing. Age, sex, and MSA information was obtained from the enrollment file. HCC variables were constructed by mapping ICD-10 diagnosis codes recorded in the inpatient, outpatient, and carrier claims using the 2019 software. We considered age categories (21-29, 30-39, 40-49, 50-59, 60-65), documented sex, and twelve chronic condition indicators (arthritis, asthma and other respiratory conditions, cancer, diabetes, heart disease, hypertension, kidney disease, hyperlipidemia, mental health and substance use disorders, nervous system conditions, osteoporosis, and viral infections) as component variables to define groups in $G$. These twelve conditions were selected because they are common chronic conditions monitored by CMS \citep{cms_chronic}. The chronic conditions indicators were created using individual and multilevel diagnosis-based categories from the Agency for Healthcare Research and Quality Clinical Classification Software (see Table A1 for mapping). These categories incorporate more ICD-10 diagnosis codes than HCCs and therefore allow for identification of health states not captured by HCCs. 

\subsection{Implementation}

We sampled 1 million individuals for each sample year and ran separate ordinary least squares regressions to predict annual spending. We then deployed the random forests algorithm to predict the residual for each sample year. We identified groups under different size and complexity hyperparameters holding the number of trees (1,000) and the number of component variables selected at each branch (10) constant. This resulted in 4 hyperparameter settings: (1) minimum size = 100 and maximum nodes = 8, (2) minimum size = 100 and maximum nodes = 64, (3) minimum size = 10,000 and maximum nodes = 8, and (4) minimum size = 10,000 and maximum nodes = 64. The random forests algorithm was run in R using the \texttt{randomForest} package \citep{liaw_classification_2002}. To ensure that identified groups persisted in all three sample years with some frequency across trees, we limited the results to groups that appeared in at least 1\% of the trees in each sample year. Predicted versus observed residuals for each identified group were compared to assess the accuracy of our predictions. 

\section{Results}

\begin{figure}
\begin{center}
\caption{Top Under- and Overcompensated Groups in the Marketplaces Risk Adjustment (minimum node size: 10,000, maximum nodes: 8)}
\includegraphics[scale=.7]{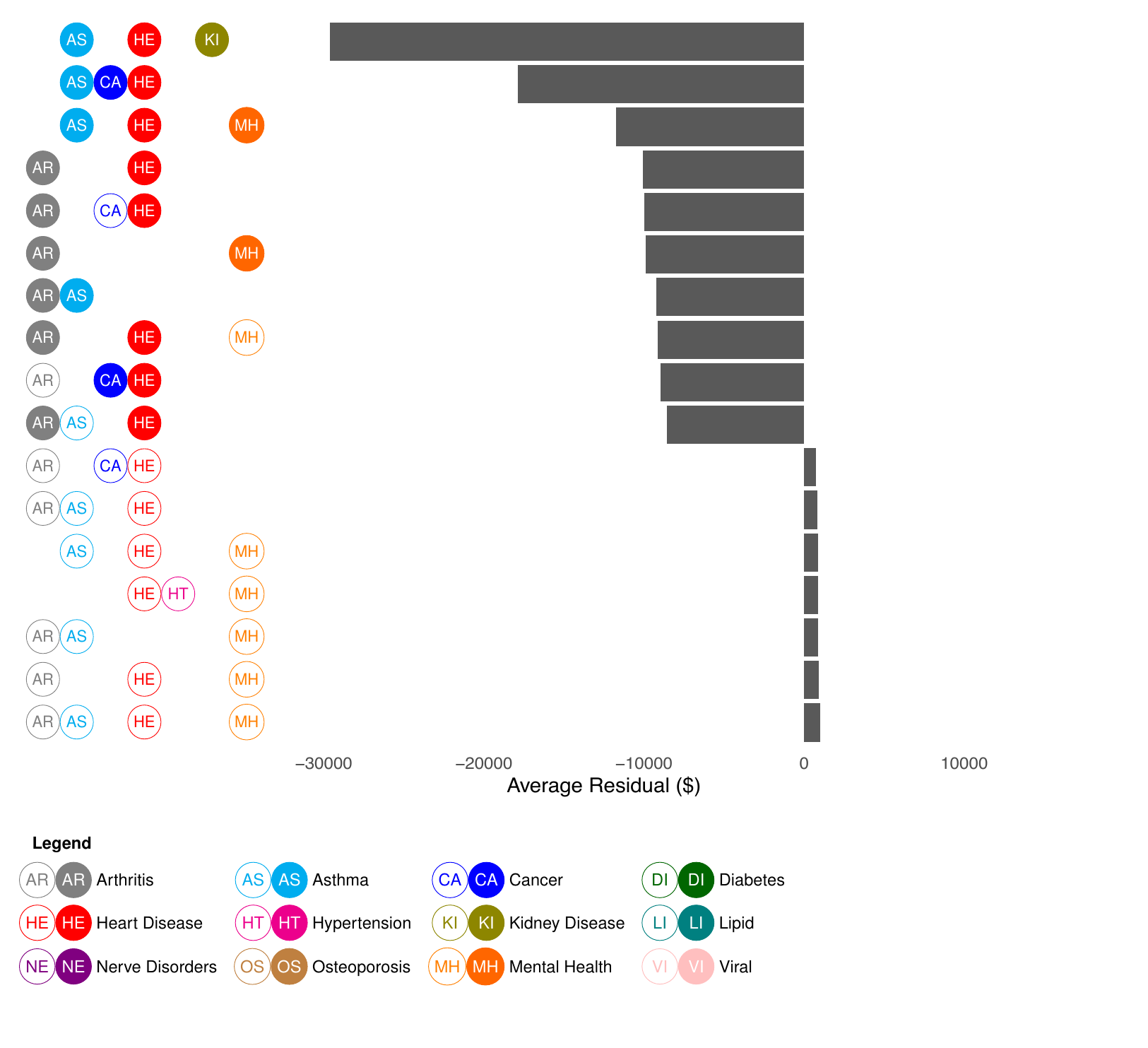} 
\end{center}
\par
\textit{Note:} Unfilled circles indicate the lack of a condition. 
Only 7 overcompensated groups were identified in this setting, 
after removing groups that appeared in less than 1\% of trees across the three years.  
\label{fig:fig1}
\end{figure}

Table 1 summarizes the Marketplaces sample characteristics for each sample year. About 5\% of the sample was documented as female. The sample was fairly evenly distributed across the five age categories with more enrollees in their 50s than any other age group (26.6 to 27.2\%). Of the 12 chronic conditions, hypertension was the most common, documented in about 14\% of the sample. Osteoporosis, chronic kidney disease, and chronic viral infections were the least common, appearing in <1\% of the sample. For each sample year, we observed an average undercompensation of around -\$1,500 to -\$5,000 for most of the selected chronic conditions. Diabetes was the least undercompensated group with an average undercompensation of a few hundred dollars (see Table 2). Average annual spending ranged from \$6,500 to \$7,000 in the study period.  

In general, all four hyperparameter settings yielded similar results. Individuals with multiple chronic conditions, in particular enrollees with some combination of asthma, heart disease, arthritis, and mental health and substance use disorders, tended to be undercompensated whereas individuals with no chronic conditions were overcompensated. In Figure 2, we present the top under- and overcompensated groups (measured by the average residual across the three sample years) limiting the group size to 10,000 and maximum nodes to 8. Predicted undercompensation was substantially larger than overcompensation. The groups we identified with multiple chronic conditions were undercompensated, on average, by at least -\$10,000 and up to -\$29,600, whereas overcompensated groups were all overcompensated by less than \$1,000.  Also, the predicted residuals for groups defined by multiple attributes were much larger than the average residuals for each condition individually. For example, individuals documented as having a combination of asthma, heart disease, and mental health and substance use disorders were estimated to be undercompensated by up to \$12,000 on average, which is more than the sum of their observed average residuals.

\begin{table}[h]
	\caption{Marketplaces Sample Characteristics (in Percents)}
	\centering
	\begin{tabular}{lrrr}
		\toprule
		\textbf{Variables}     & \textbf{2016}     & \textbf{2017} & \textbf{2018} \\
		\midrule
		Documented Sex, Female & 52.3 & 52.2 & 51.6  \\
		Age \\
		\quad 21-29 & 17.9	& 17.9	& 18.0 \\
		\quad 30-39 & 20.4	& 20.6	& 21.0 \\
		\quad 40-49 & 23.7	& 23.7	& 23.7  \\
		\quad 50-59 & 27.2	& 26.9	& 26.6 \\
		\quad 60-65 & 10.8	& 10.8	& 10.7  \\
Arthritis & 4.5	& 4.5	& 4.5 \\
Asthma & 	10.6	& 10.7	& 10.8 \\
Cancer	& 7.1	& 7.0	& 6.8 \\
Diabetes	& 8.9	& 9.0	& 8.9 \\
Heart Disease	& 9.1	& 9.1	& 9.3 \\
Hypertension	& 14.1	& 13.9	& 13.7 \\
Kidney Disease & 	0.6	& 0.6	& 0.6 \\
Lipid Disorders	 & 10.2 & 	9.7 & 	9.5 \\
Mental Health	& 11.1	& 11.7	& 12.6 \\
Nervous	& 0.7	& 0.7	& 0.7 \\
Osteoporosis	& 0.6	& 0.6	& 0.6 \\
Viral Infections	& 0.4	& 0.4	& 0.4 \\
		\bottomrule 
		N & 1,000,000 & 1,000,000 & 1,000,000 \\
	\end{tabular}
	\label{tab:table}
\end{table}

\begin{table}[h]
	\caption{Marketplaces Average Residual (\$) by Chronic Condition}
	\centering
	\begin{tabular}{lrrrrrr}
		\toprule
		& \multicolumn{ 2}{c}{\textbf{2016} }    & \multicolumn{ 2}{c}{\textbf{2017} }  & \multicolumn{ 2}{c}{\textbf{2018} } \\
		\textbf{Chronic Condition}     & No & Yes & No & Yes & No & Yes \\
		\midrule
Arthritis	& 263	& -5611 & 	293	& -6213	& 283	& -5957 \\
Asthma	 & 325	& -2751	& 335	& -2783	& 458	& -3785 \\
Cancer	& 163	& -2120	& 165	& -2200	& 189	& -2583 \\
Diabetes	& 23	& -236	& 19	 & -188	& 47	& -481 \\
Heart Disease &	393	 & -3924	 & 383	 & -3813	 & 502	 & -4903 \\
Hypertension &	263	 & -1596	 & 255	 & -1583	 & 315	 & -1985 \\
Kidney Disease &	18	 & -2838	 & 20	 & -3126	 & 28	 & -4596 \\
Lipid Disorders	 &121	 & -1065	 & 113	 & -1044	 & 98	 & -930 \\
Mental Health &	312	 & -2499	 & 330	 & -2492	 & 398	 &  -2770 \\
Nervous &	33	 & -4775	 & 30	 & -4382	 & 36	 & -5229 \\
Osteoporosis &	12	 & -1963	 & 15	 & -2593	 & 10	 & -1760 \\
Viral Infections	 & 2	 & -419	 & 6	 & -1508	 & 5	 & -1276 \\
	\end{tabular}
	\label{tab:table}
\end{table}

When we allowed for smaller (minimum group size = 100) and more complex groups (maximum nodes = 64) we saw groups with more conditions appear in the top ten (Figure 3). Namely, we identified more groups with asthma, cancer, hypertension, chronic kidney disease, and mental health and substance use disorders. Interestingly, we also found that most top overcompensated groups were older, but relatively healthy individuals.  In general, tuning the maximum node size, which limits the depth of the trees, was more important for determining group complexity than limiting the size of the groups. Some chronic conditions, such as chronic viral infections and disorders of lipid metabolism as well as documented sex, rarely if ever appeared as components in our identified groups.

When we disaggregated the results by year, we observed the same pattern of undercompensation for enrollees with multiple chronic conditions and overcompensations for those without, but there was some temporal variation in the types of conditions that were identified (Figure A1). For example, in 2016, five out of the ten top undercompensated groups were identified as having a hereditary and generative nervous system in addition to other chronic conditions, but this attribute didn’t appear in later years. 
	
In the Appendix, we present results from the other setting we considered: identifying undercompensated groups in the Medicare risk adjustment formula used by CMS to set capitation rates for Medicare Advantage plans. We found no groups that were consistently under- or overcompensated across the three sample years. 

\begin{figure}
\begin{center}
\caption{Top Under- and Overcompensated Groups in the Marketplaces Risk Adjustment (minimum node size: 100, maximum nodes: 64)}
\includegraphics[scale=.7]{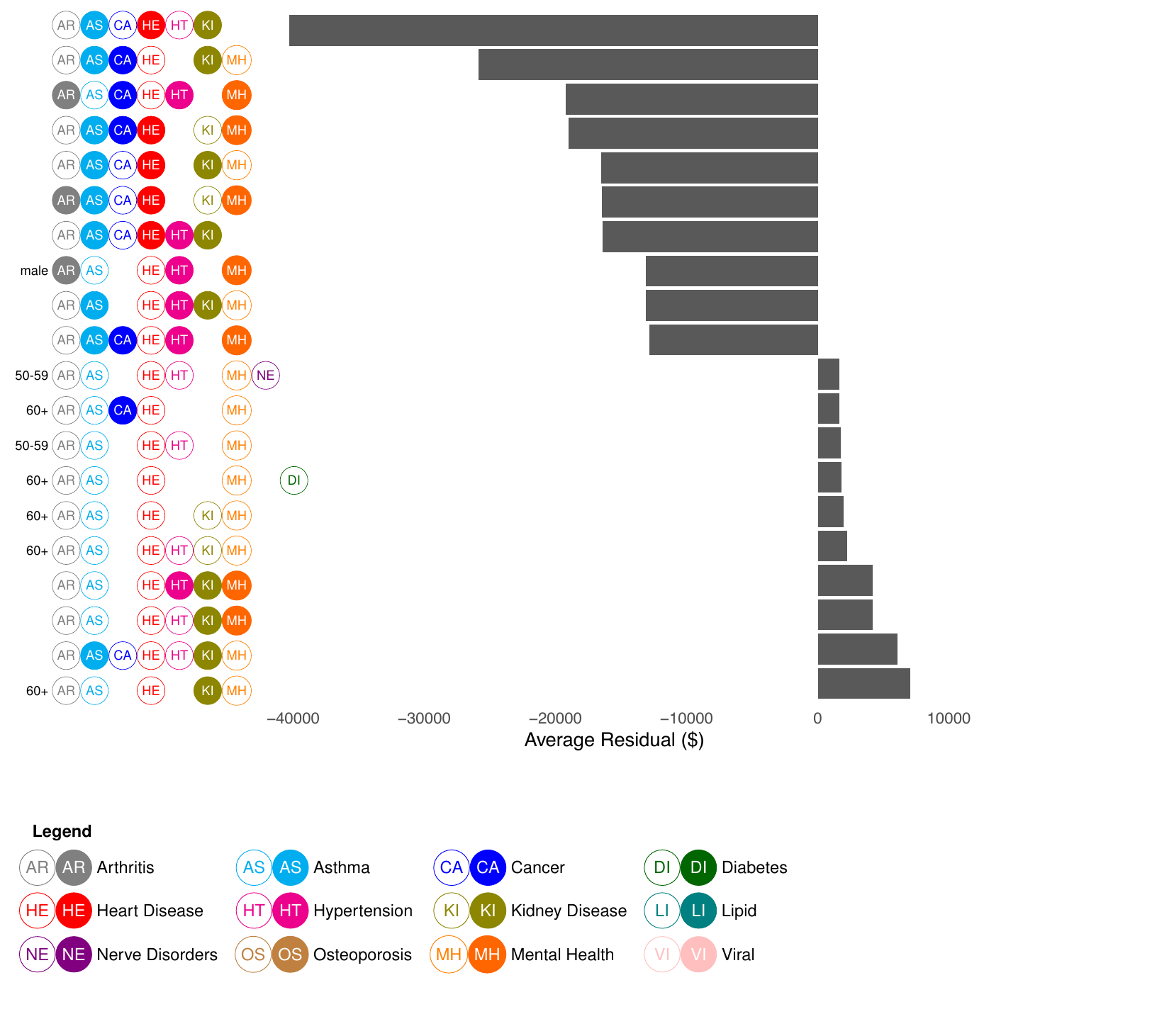}
\end{center}
\textit{Note:} Unfilled circles indicate the lack of a condition. 
\label{fig:fig1}
\end{figure}

\section{Discussion}

In this research, we presented a new approach for identifying undercompensated groups defined by multiple attributes, extending the random forests algorithm to determine group importance without requiring groups to be pre-specified. We implemented our group importance algorithm on the Marketplaces and Medicare risk adjustment formulas. The Marketplaces formula favored healthier individuals compared to those with multiple chronic conditions. In particular, groups that had some combination of asthma, heart disease, arthritis, and mental health and substance use disorder were largely undercompensated. These results expand on and reenforce earlier work exploring issues of undercompensation in risk adjustment for people with multiple chronic conditions \cite{pope_evaluation_2011}. Undercompensation in the Medicare risk adjustment formula was smaller in magnitude compared to the Marketplaces formula and no groups were persistently under- or overcompensated across the three sample years. This suggests that the Medicare risk adjustment formula may induce fewer incentives to discriminate with respect to the twelve selected conditions. 

Our results raise the question of how to address incentives in markets, especially for people with multiple health conditions. There are numerous approaches for improving risk adjustment performance aimed at removing selection incentives in regulated insurance markets. The Marketplaces aim to partly address undercompensation for enrollees with two chronic health conditions, where one condition is severe, by including interactions in the risk adjustment formula \citep{kautter_hhs-hcc_2014}. Given our results, additional interactions for those with multiple chronic conditions could be beneficial.  However, in Germany and the Netherlands, researchers found mixed results when they included the interactions found in a single regression tree predicting the residual, sometimes leading to negative consequences for other groups \citep{buchner_regression_2017, van_veen_exploring_2018}. While not yet used in practice, machine learning methods have also been proposed as an alternative to predict spending and may be better able to capture non-linear spending trends for enrollees with multiple chronic conditions \citep{rose_machine_2016}. Another plausible option could be to enforce improved fit for selected groups through constrained or penalized regression, incorporating group fairness directly into the loss function and where negative consequences for other groups were not observed \citep{zink_fair_2020, mcguire_simplifying_2021}. Although, solutions to discrimination via insurance benefit design may lie outside of the risk adjustment formula itself and be better addressed by, for example, additional legal remedies, such as requiring adequate coverage of treatments and services by insurers for the set of conditions we identified. 

This work has several limitations. We developed a methodology that focused on groups with the largest magnitude average residual that appeared in all three sample years. The groups were composed of conditions that we considered actionable for insurers, but we didn’t consider a priori whether certain combinations of conditions (or lack of conditions) would be more actionable for insurers compared to others. In our algorithm, we set the size and number of terminal nodes in order to restrict the size and complexity of a group, but future work could develop more sophisticated algorithms that aim to empirically examine whether specific combinations of chronic conditions are actionable for insurers within the decision rules of the tree rather than post hoc. 

The techniques we presented are relevant to other insurance markets relying on risk adjustment. They can be used by developers and evaluators of risk adjustment formulas to identify and assess performance across many different groups defined by multiple attributes. In this paper, we focused on twelve chronic conditions, but our method allows for the inclusion of additional chronic conditions and other variables appropriate for identifying groups, subject to computational constraints and ethical factors. Variable choices will be limited by data availability: while countries like the U.S. and the Netherlands have rich data sources to construct and assess their risk adjustment formulas, not all countries have access to such detailed data \citep{mcguire_risk_2018}.

Our group importance method may be relevant in a wide range of applications beyond risk adjustment, although we caution researchers to carefully consider the context before attempting to identify potentially marginalized groups. In some settings, identifying groups could actively cause harm, e.g., may involve collecting or amplifying stigmatizing information. In other cases, our tool may be useful in mitigating ongoing harms. Machine learning predictions for clinical outcomes have been found to be less accurate for groups defined by age, race, or other attributes, contributing to health disparities \cite{chen_ethical_2021}. Our new method could help ensure that algorithms deployed in such settings remedy inequities for currently unidentified marginalized groups. We recommend researchers create a social impact statement and follow an ethical pipeline for building algorithms when considering adapting our tool to any setting \citep{chen_ethical_2021, principles}. 


\bibliographystyle{asa}
\bibliography{references}  

\newpage

\appendix

\renewcommand{\thefigure}{A\arabic{figure}}
\renewcommand{\thetable}{A\arabic{table}}

\section{Appendix}

\subsection{Medicare Risk Adjustment}

The Medicare Risk Adjustment formula was first implemented in 2004 by Centers for Medicare \& Medicaid Services (CMS) to set capitation rates for Medicare Advantage plans and ensure that plans with higher risk beneficiaries (on average) are compensated more than plans with lower risks beneficiaries (on average). The formula is built and calibrated using Traditional Medicare beneficiaries. Our risk adjustment estimates were based on the 2017 Medicare risk adjustment formula for beneficiaries who qualified for Medicare based on age and were not enrolled in Medicaid. We additionally included a control for county, the level at which premiums are set: 

$$Y=\textnormal{Age} \times \textnormal{Sex}+\textnormal{HCC}+\textnormal{County}+\epsilon,$$

where Age $\times$ Sex are eight age and sex categories and HCC are a set of Health Condition Categories. Unlike the Marketplaces formulas, the Medicare risk adjustment formula is prospective, meaning it uses diagnosis-based condition categories from the previous year to predict current year spending. 

\subsubsection{Data}

To evaluate the Medicare risk adjustment formula, we sampled individuals age 65+ enrolled in Traditional Medicare for at least two years starting in either 2015, 2016, or 2017. Because the risk adjustment formula is prospective, we required these two years of enrollment in order to have a complete year of data for measuring HCCs and a subsequent complete year of data for annual spending. Individual annual spending, measured in 2016, 2017, and 2018, was calculated as the total spending across the individual's outpatient, inpatient, and carrier claims in the given year. Diagnosis codes recorded in inpatient, outpatient, and carrier claims were mapped to HCC categories using the CMS-HCC software. The component variables in $I$ to define groups in $G$ were age categories (65-69, 70-79, 80-89, 90+), sex, and the same four chronic conditions (mental health disorders, cancer, heart disease, and diabetes). Similar to the Marketplaces formula, chronic conditions were defined using the individual diagnosis-based categories from the Agency for Healthcare Research and Quality Clinical Classification Software but measured in the prior year. Age and sex information was acquired from the Master Beneficiary Summary File. We excluded individuals with any months of Medicaid eligibility in their starting year.

\subsubsection{Results}

The Medicare sample had more individuals with female documented sex than the Marketplaces sample (55\% compared to 52\%), and nearly half the sample was in their 70s (45.1\% to 47.7\%). As expected for an older sample, chronic conditions were much more prevalent than in the younger commercial sample: a majority of the conditions were found in at least 25\% of the sample (hypertension and lipid disorders were the most common conditions with a prevalence of approximately 65\%).  Chronic viral infections and nervous conditions were the rarest conditions and were only recorded in 0.4\% and 6.8\% of the sample, respectively. Recall that chronic conditions were recorded in the year prior, and the slight differences in prevalence for the 2016 sample (with 2015 diagnosis) may be due to the switch from ICD-9 to ICD-10 on October 1, 2015. Table A2 presents the sample statistics for the Medicare sample. The observed average residual for every chronic condition except chronic viral infections was trivial (all less than \$100, or less than 1\% of the average annual spending of \$7,500). Chronic viral infections had an average undercompensation of -\$358 in 2016 and average overcompensation in both 2017 and 2018 (\$138 and \$184, respectively). In general, under- versus overcompensation did not appear uniformly across years (Table A3).
	We found no groups that were consistently under- and overcompensated over the three-year observation period. In Figure A6, we display the top under- and overcompensated groups for each sample year when limiting the group size to 10,000 and maximum nodes to 8. The top identified groups are different in each year, with certain conditions appearing in some years and not others. For example, in 2017 groups with chronic kidney disease appeared in six out of the ten top undercompensated groups, but not consistently in other years. Some conditions appeared in both under- and overcompensated groups within and across years. For example, nerve disorders was a common condition in groups that were under- and overcompensated in 2017 and 2018. Six out of the ten top undercompensated groups in 2016 included beneficiaries with viral infections, but groups with this condition tended to be overcompensated in 2017 and 2018. Recall that chronic viral infections had the largest observed negative average residual in 2016 and largest positive residual in 2017 and 2018, thus, this result aligns with those findings. Compared to the Marketplaces risk adjustment, the magnitude of the predicted average residual for undercompensation was smaller (about \$500 rather than thousands) and more evenly balanced between the under- and overcompensated groups. When we allowed the groups to be more complex and smaller in size, we identified very few groups that appeared in more than 1\% of trees.

\setcounter{table}{0}    

\begin{table} 
\caption{Mapping Diagnosis Codes and Clinical Classification Software (CCS) to Chronic Conditions}
	\centering
	\begin{tabular}{llccc}
		\toprule
		&      & \textbf{Single Level} & \textbf{Multiple 2nd Level} & \textbf{ICD-10 } \\
				   &  & \textbf{CCS Diagnosis} & \textbf{CCS Diagnosis} &  \textbf{Diagnosis} \\
		 \textbf{Chronic Condition}    & \textbf{Short Name}  & \textbf{Categories} & \textbf{Categories} & \textbf{Codes} \\
		 \hline 
Arthritis (osteoarthritis \& rheumatoid)	 & Arthritis	& 202,203	&  \\	
Asthma, COPD, and Other Chronic Lung Diseases	&  Asthma	&  & 8.2-8.8	 \\
Cancer	& Cancer	& 11-45		\\
Diabetes	& Diabetes	& 49, 50		 \\
Diseases of the Heart	& Heart	 & 103		\\
Hypertension	& Hypertension	& 98,99		\\
Chronic Kidney Disease	& Kidney	& 158		\\
Disorders of Lipid Metabolism	& Lipid	& 53		\\
Mental Health and Substance Use Disorders	& Mental	& 650-663, 670 \\		
Hereditary and Degenerative Nervous System	& Nervous	& & 	6.2	\\
Osteoporosis	& Osteoporosis	& 206		\\
Chronic Viral Infections & Viral	&  5 &		& B18		   \\
			\end{tabular}
	\label{tab:table}
\end{table}

\begin{table}
	\caption{Medicare Sample Characteristics (in Percents)}
	\centering
	\begin{tabular}{lrrr}
		\toprule
		\textbf{Variables}     & \textbf{2016}     & \textbf{2017} & \textbf{2018} \\
		\midrule
		Documented Sex, Female & 55.2	& 55.1	& 55.0 \\
		Age \\
		\quad 65-69 & 27.9	& 27.6	& 26.6\\
		\quad 70-79 & 45.1	& 46.2	& 47.6 \\
		\quad 80-89 & 22.1	 & 21.3	& 21.0  \\
		\quad 90+ & 4.9	& 4.8	& 4.7 \\
Arthritis	 & 28.9	 & 30.1	 & 30.7 \\
Asthma	 & 37.4	 & 37.4	 & 38.8 \\
Cancer	 & 33.0	 & 33.3	 & 33.9 \\
Diabetes	 & 35.7	 & 37.2	 & 38.3 \\
Heart Disease	 & 45.6	 & 45.3	 & 45.8 \\
Hypertension	 & 68.6	 & 68.5	 & 68.3 \\
Kidney Disease & 	11.6	 & 12.2	 & 12.8 \\
Lipid Disorders & 	65.7	 & 65.4	 & 66.0 \\
Mental Health	 & 34.2	 & 35.6 & 	37.7 \\
Nervous	 & 6.8	 & 7.3	 & 7.5 \\
Osteoporosis	 & 11.1	 & 10.8	 & 11.0 \\
Viral Infections & 	0.4	 & 0.4	 & 0.4 \\
		\bottomrule 
		N & 1,000,000 & 1,000,000 & 1,000,000 \\
	\end{tabular}
	\label{tab:table}
\end{table}
	
	\begin{table}
	\caption{Medicare Average Residual (\$) by Chronic Condition}
	\centering
	\begin{tabular}{lrrrrrr}
		\toprule
		& \multicolumn{ 2}{c}{\textbf{2016} }    & \multicolumn{ 2}{c}{\textbf{2017} }  & \multicolumn{ 2}{c}{\textbf{2018} } \\
		\textbf{Chronic Condition}     & No & Yes & No & Yes & No & Yes \\
		\midrule
Arthritis	 & -19	 & 47 & 	6	 & -14	 & 4	 & -8 \\
Asthma	  & 6	 & -10	 & -6	 & 10	 & 8	 & -13 \\
Cancer	 & 17	 & -34	 & -11	 & 23	 & -1 & 	3 \\
Diabetes & 	-11	 & 20	 & -8	 & 13	 & 9	 & -14 \\
Heart Disease	 & 2	 & -2	 & 12	 & -14	 & 10	 & -11 \\
Hypertension	 & 12	 & -5	 & 5	 & -2	 & 22	 & -10 \\
Kidney Disease	 & -5	 & 40	 & 11	 & -80	 & 3	 & -21 \\
Lipid Disorders	 & -16	 & 8	 & -5	 & 3	 & -8	 & 4 \\
Mental Health	 & 10	 & -20	 & -8	 & 15	 & 7	 & -12 \\
Nervous	 & -1	 & 20	 & 2	 & -26	 & -4	 & 46 \\
Osteoporosis	 & 0	 & 3	 & -12	 & 96	 & -1	 & 10 \\
Viral Infections	 & 1	 & -358	 & 0	 & 138	 & 0	 & 184 \\
	\end{tabular}
	\label{tab:table}
\end{table}
	
\newpage 

\setcounter{figure}{0}    

\begin{figure}
\begin{center}
\caption{Top Under- and Overcompensated Groups in the Marketplaces Risk Adjustment by Year (minimum node size: 10,000, maximum nodes: 8)}
\includegraphics[scale=1]{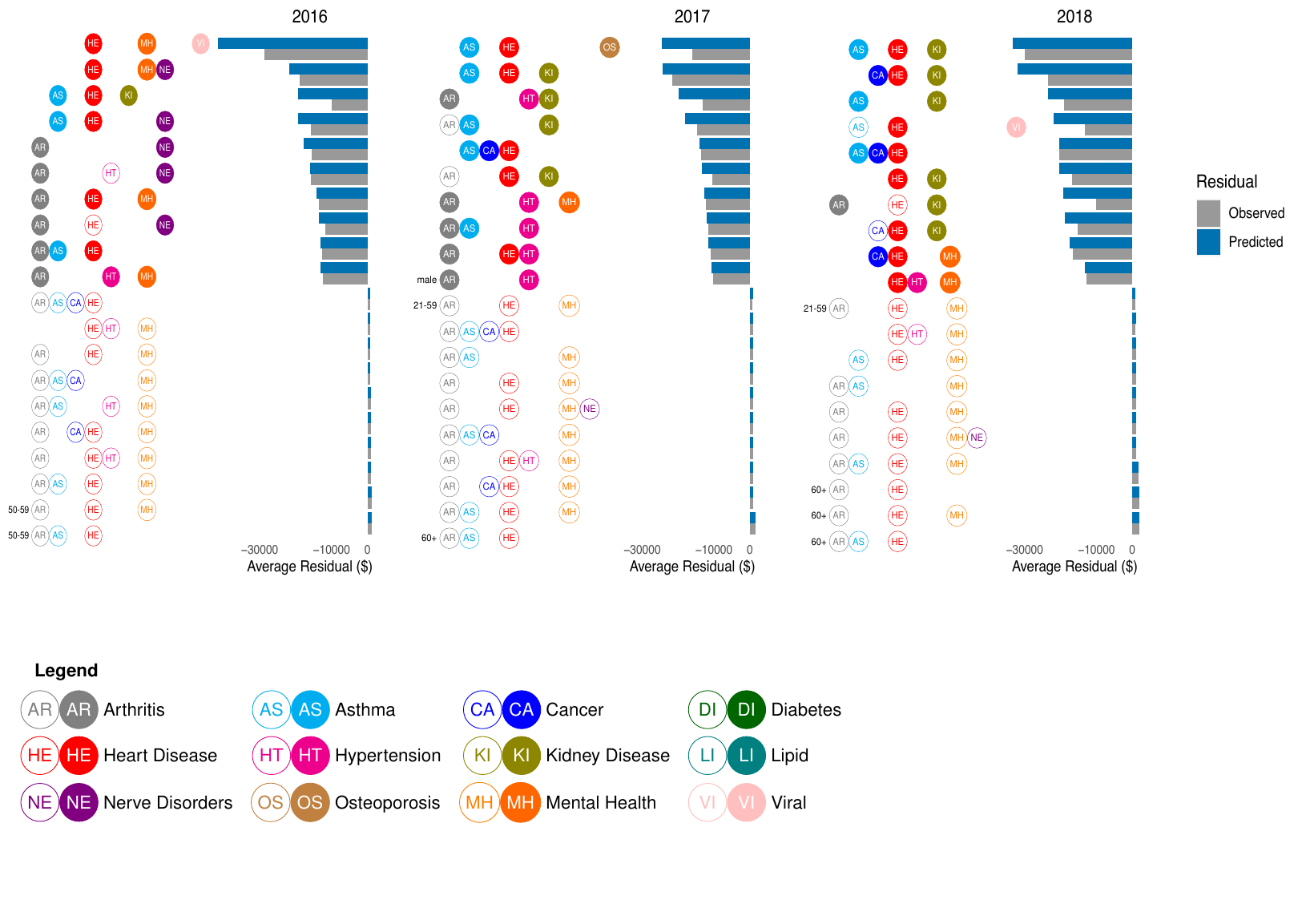}
\end{center}
\textit{Note:} Unfilled circles indicate the lack of a condition. 
\label{fig:figS1}
\end{figure}

\newpage

\begin{figure}
\begin{center}
\caption{Top Under- and Overcompensated Groups in the Marketplaces Risk Adjustment, Observed vs Predicted Residuals (minimum node size: 10,000, maximum nodes: 8) }
\includegraphics[scale=1]{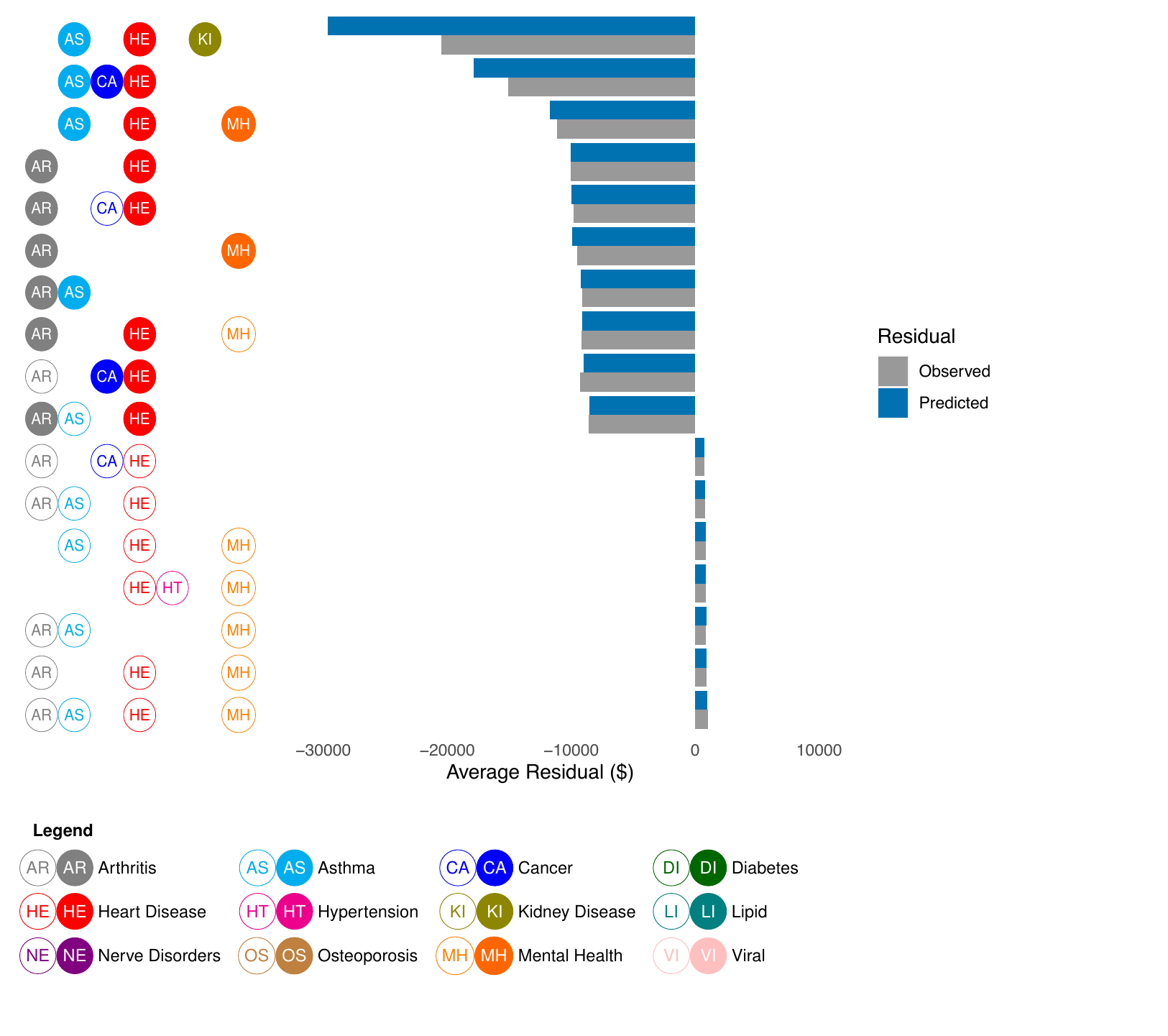}
\end{center}
\textit{Note:} Unfilled circles indicate the lack of a condition. 
\label{fig:figS2}
\end{figure}

\newpage

\begin{figure}
\begin{center}
\caption{Top Under- and Overcompensated Groups in the Marketplaces Risk Adjustment, Observed vs Predicted Residuals (minimum node size: 100, maximum nodes: 64) }
\includegraphics[scale=1]{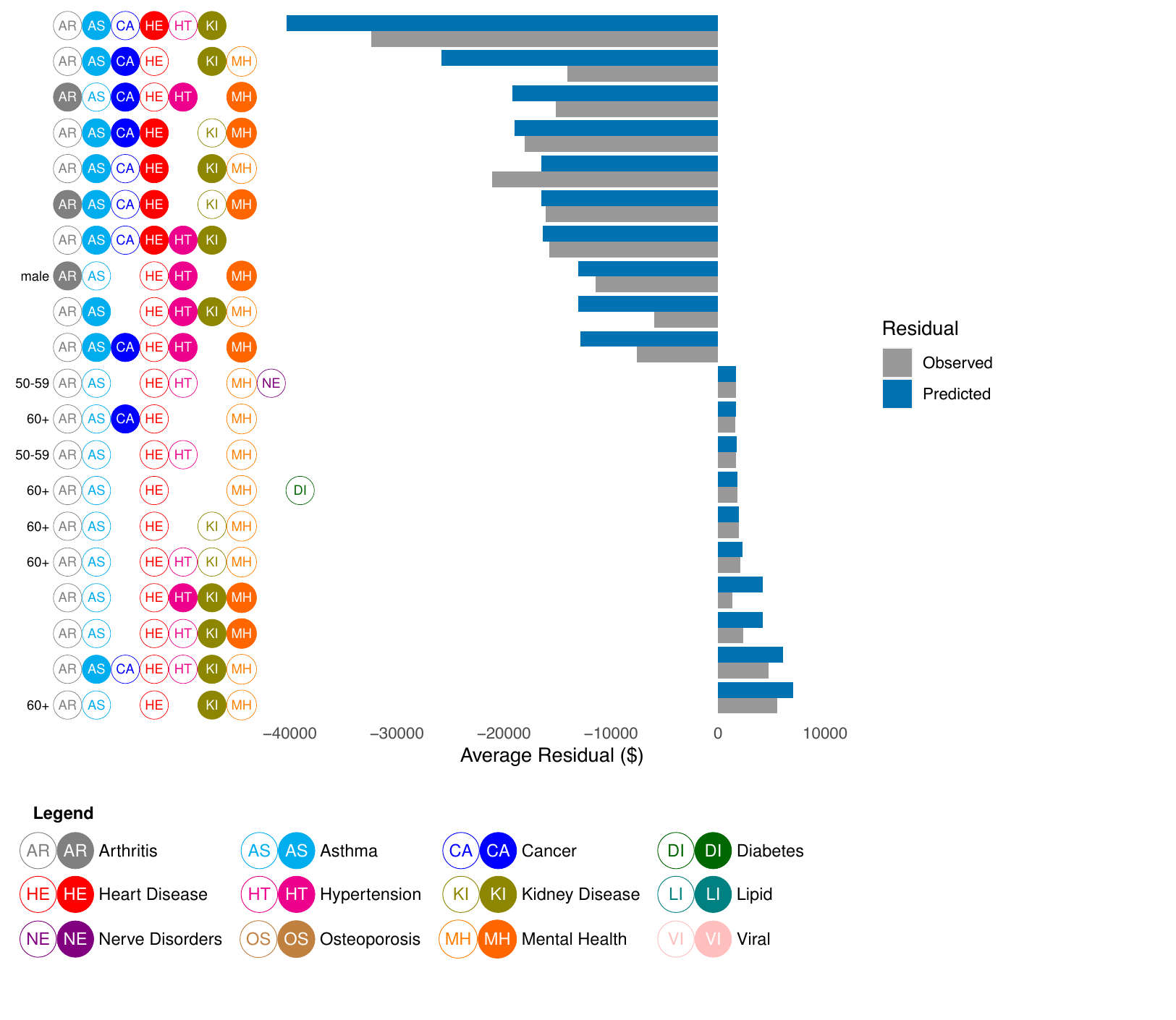}
\end{center}
\textit{Note:} Unfilled circles indicate the lack of a condition. 
\label{fig:figS3}
\end{figure}

\newpage

\begin{figure}
\begin{center}
\caption{Top Under- and Overcompensated Groups in the Marketplaces Risk Adjustment, Observed vs Predicted Residuals (minimum node size: 10,000, maximum nodes: 64)}
\includegraphics[scale=1]{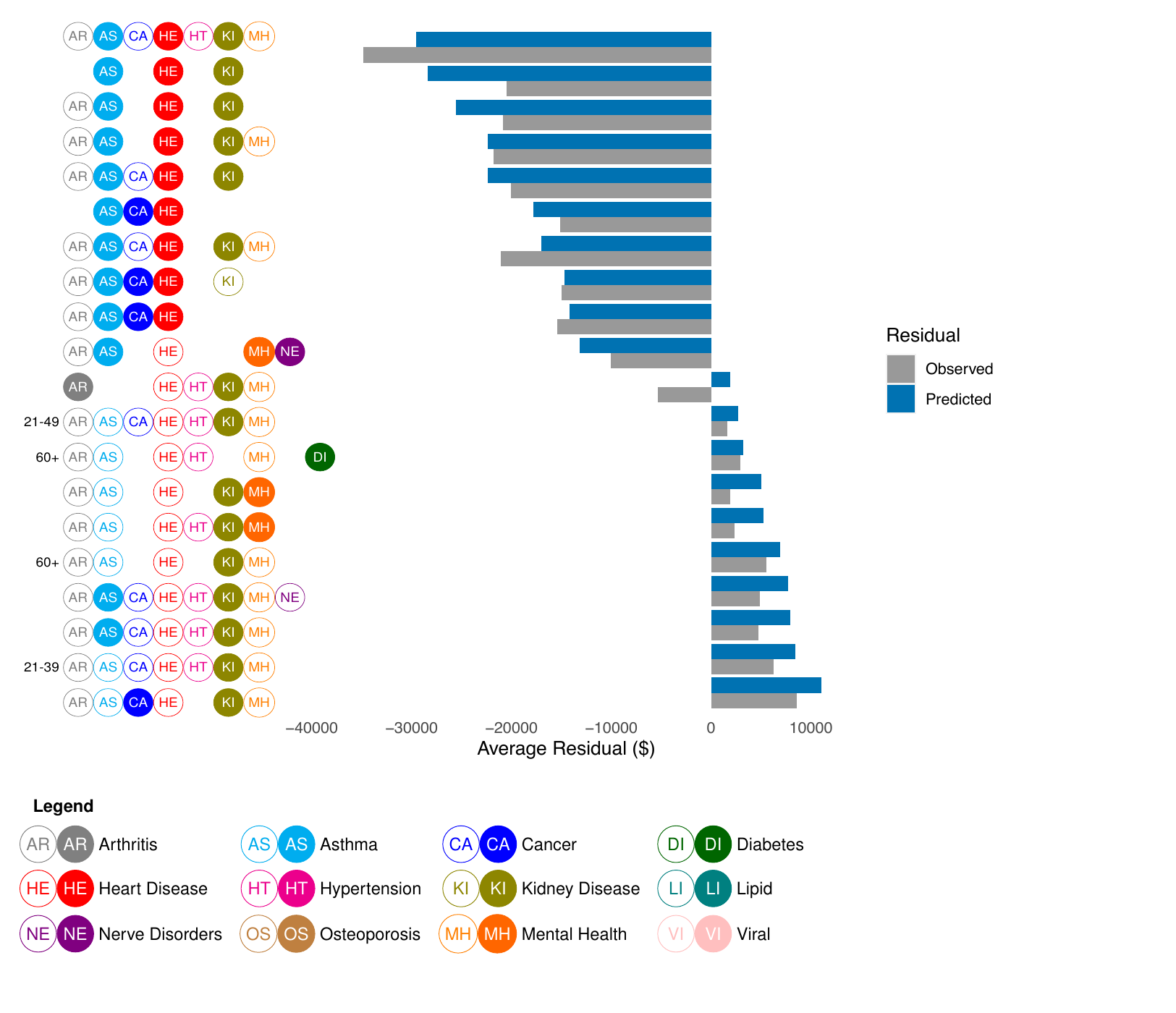}
\end{center}
\textit{Note:} Unfilled circles indicate the lack of a condition. 
\label{fig:figS4}
\end{figure}

\newpage

\begin{figure}
\begin{center}
\caption{Top Under- and Overcompensated Groups in the Marketplaces Risk Adjustment, Observed vs Predicted Residuals (minimum node size: 100, maximum nodes: 8)}
\includegraphics[scale=1]{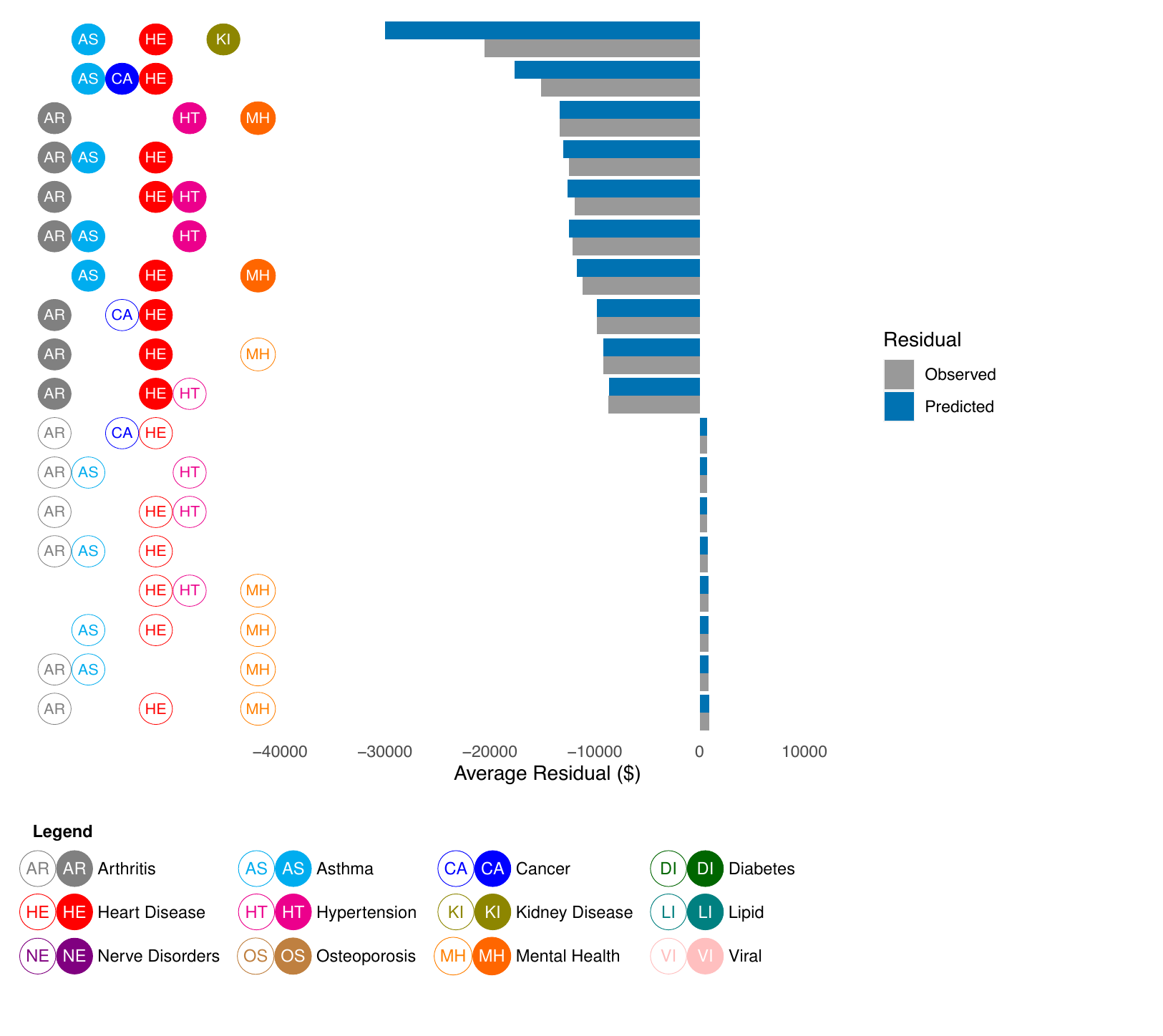}
\end{center}
\textit{Note:} Unfilled circles indicate the lack of a condition. 
\label{fig:figS5}
\end{figure}

\clearpage

\begin{figure}
\begin{center}
\caption{Top Under- and Overcompensated Groups in the Medicare Risk Adjustment by Year (minimum node size: 10,000, maximum nodes: 8)}
\includegraphics[scale=1]{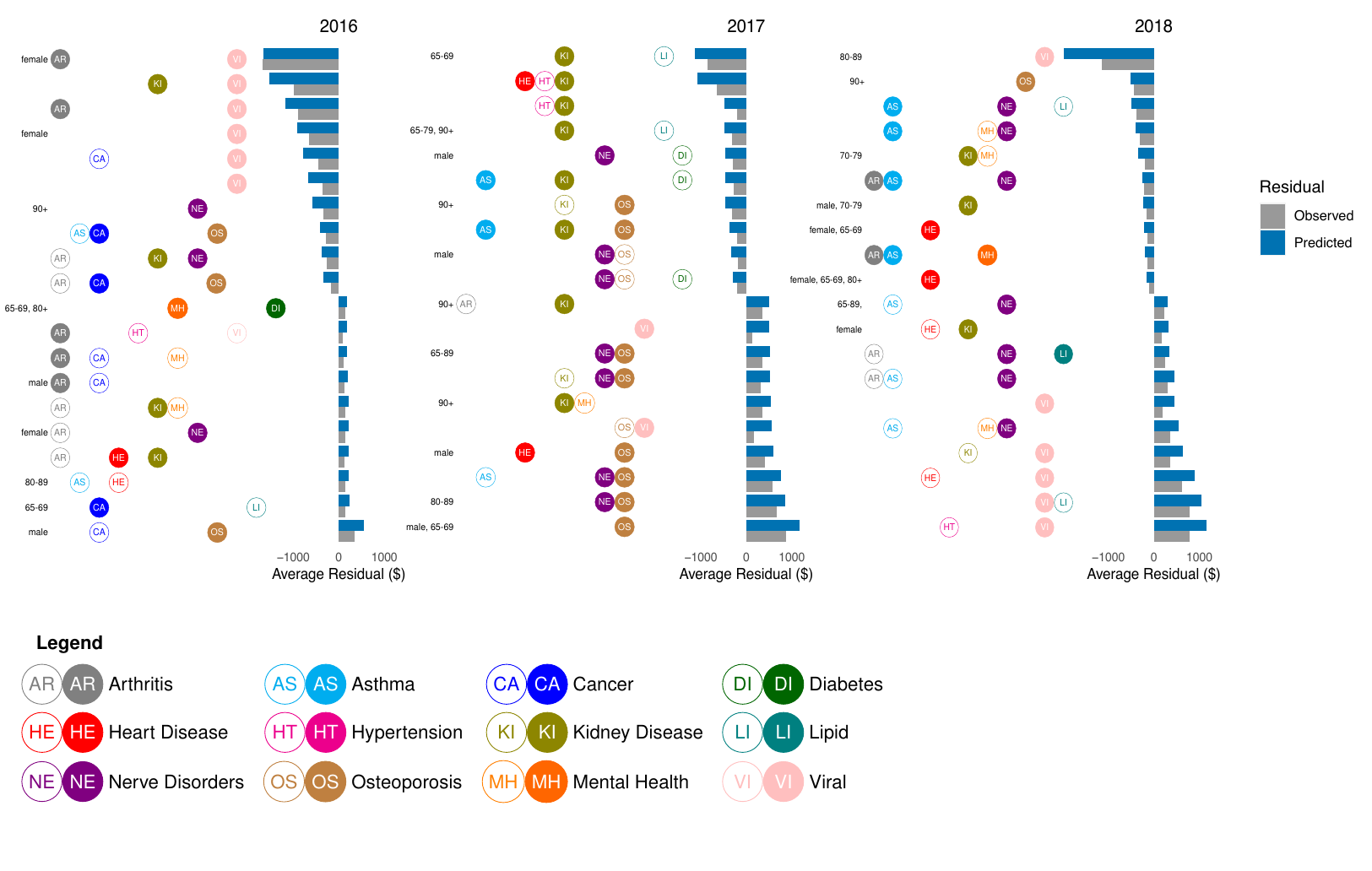}
\end{center}
\textit{Note:} Unfilled circles indicate the lack of a condition. 
\label{fig:figS6}
\end{figure}

\end{document}